\documentstyle[twoside]{article}

\catcode`\@=11
\long\def\@makefntext#1{
\protect\noindent \hbox to 3.2pt {\hskip-.9pt  
$^{{\eightrm\@thefnmark}}$\hfil}#1\hfill}		

\def\@makefnmark{\hbox to 0pt{$^{\@thefnmark}$\hss}}	
	
\def\ps@myheadings{\let\@mkboth\@gobbletwo
\def\@oddhead{\hbox{}
\rightmark\hfil\eightrm\thepage}   
\def\@oddfoot{}\def\@evenhead{\eightrm\thepage\hfil
\leftmark\hbox{}}\def\@evenfoot{}
\def\sectionmark##1{}\def\subsectionmark##1{}}

\oddsidemargin=\evensidemargin
\newcounter{sectionc}\newcounter{subsectionc}\newcounter{subsubsectionc}
\renewcommand{\section}[1] {\vspace{12pt}\addtocounter{sectionc}{1} 
\setcounter{subsectionc}{0}\setcounter{subsubsectionc}{0}\noindent 
	{\tenbf\thesectionc. #1}\par\vspace{5pt}}
\renewcommand{\subsection}[1] {\vspace{12pt}\addtocounter{subsectionc}{1} 
	\setcounter{subsubsectionc}{0}\noindent 
	{\bf\thesectionc.\thesubsectionc. {\kern1pt \bfit #1}}\par\vspace{5pt}}
\renewcommand{\subsubsection}[1] {\vspace{12pt}\addtocounter{subsubsectionc}{1}
	\noindent{\tenrm\thesectionc.\thesubsectionc.\thesubsubsectionc.
	{\kern1pt \tenit #1}}\par\vspace{5pt}}
\newcommand{\nonumsection}[1] {\vspace{12pt}\noindent{\tenbf #1}
	\par\vspace{5pt}}

\newcounter{appendixc}
\newcounter{subappendixc}[appendixc]
\newcounter{subsubappendixc}[subappendixc]
\renewcommand{\thesubappendixc}{\Alph{appendixc}.\arabic{subappendixc}}
\renewcommand{\thesubsubappendixc}
	{\Alph{appendixc}.\arabic{subappendixc}.\arabic{subsubappendixc}}

\renewcommand{\appendix}[1] {\vspace{12pt}
        \refstepcounter{appendixc}
        \setcounter{figure}{0}
        \setcounter{table}{0}
        \setcounter{lemma}{0}
        \setcounter{theorem}{0}
        \setcounter{corollary}{0}
        \setcounter{definition}{0}
        \setcounter{equation}{0}
        \renewcommand{\thefigure}{\Alph{appendixc}.\arabic{figure}}
        \renewcommand{\thetable}{\Alph{appendixc}.\arabic{table}}
        \renewcommand{\theappendixc}{\Alph{appendixc}}
        \renewcommand{\thelemma}{\Alph{appendixc}.\arabic{lemma}}
        \renewcommand{\thetheorem}{\Alph{appendixc}.\arabic{theorem}}
        \renewcommand{\thedefinition}{\Alph{appendixc}.\arabic{definition}}
        \renewcommand{\thecorollary}{\Alph{appendixc}.\arabic{corollary}}
        \renewcommand{\theequation}{\Alph{appendixc}.\arabic{equation}}
        \noindent{\tenbf Appendix \theappendixc #1}\par\vspace{5pt}}
\newcommand{\subappendix}[1] {\vspace{12pt}
        \refstepcounter{subappendixc}
        \noindent{\bf Appendix \thesubappendixc. {\kern1pt \bfit #1}}
	\par\vspace{5pt}}
\newcommand{\subsubappendix}[1] {\vspace{12pt}
        \refstepcounter{subsubappendixc}
        \noindent{\rm Appendix \thesubsubappendixc. {\kern1pt \tenit #1}}
	\par\vspace{5pt}}

\newcommand{\textlineskip}{\baselineskip=13pt}
\newcommand{\smalllineskip}{\baselineskip=10pt}

\def\eightcirc{
\begin{picture}(0,0)
\put(4.4,1.8){\circle{6.5}}
\end{picture}}
\def\eightcopyright{\eightcirc\kern2.7pt\hbox{\eightrm c}} 

\newcommand{\copyrightheading}[1]
	{\vspace*{-2.5cm}\smalllineskip{\flushleft
	{\footnotesize International Journal of Modern Physics A, #1}\\
	{\footnotesize World Scientific Publishing
	 Company}\\
	 }}

\def\abstracts#1#2#3{{
	\centering{\begin{minipage}{4.5in}\baselineskip=10pt\footnotesize
	\parindent=0pt #1\par 
	\parindent=15pt #2\par
	\parindent=15pt #3
	\end{minipage}}\par}} 

\renewenvironment{thebibliography}[1]
	{\frenchspacing
	 \ninerm\baselineskip=11pt
	 \begin{list}{\arabic{enumi}.}
	{\usecounter{enumi}\setlength{\parsep}{0pt}
	 \setlength{\leftmargin 12.7pt}{\rightmargin 0pt} 
	 \setlength{\itemsep}{0pt} \settowidth
	{\labelwidth}{#1.}\sloppy}}{\end{list}}

\newcounter{itemlistc}
\newcounter{romanlistc}
\newcounter{alphlistc}
\newcounter{arabiclistc}

\newcommand{\fcaption}[1]{
        \refstepcounter{figure}
        \setbox\@tempboxa = \hbox{\footnotesize Fig.~\thefigure. #1}
        \ifdim \wd\@tempboxa > 5in
           {\begin{center}
        \parbox{5in}{\footnotesize\smalllineskip Fig.~\thefigure. #1}
            \end{center}}
        \else
             {\begin{center}
             {\footnotesize Fig.~\thefigure. #1}
              \end{center}}
        \fi}

\newcommand{\tcaption}[1]{
        \refstepcounter{table}
        \setbox\@tempboxa = \hbox{\footnotesize Table~\thetable. #1}
        \ifdim \wd\@tempboxa > 5in
           {\begin{center}
        \parbox{5in}{\footnotesize\smalllineskip Table~\thetable. #1}
            \end{center}}
        \else
             {\begin{center}
             {\footnotesize Table~\thetable. #1}
              \end{center}}
        \fi}

\def\@citex[#1]#2{\if@filesw\immediate\write\@auxout
	{\string\citation{#2}}\fi
\def\@citea{}\@cite{\@for\@citeb:=#2\do
	{\@citea\def\@citea{,}\@ifundefined
	{b@\@citeb}{{\bf ?}\@warning
	{Citation `\@citeb' on page \thepage \space undefined}}
	{\csname b@\@citeb\endcsname}}}{#1}}

\newif\if@cghi
\def\cite{\@cghitrue\@ifnextchar [{\@tempswatrue
	\@citex}{\@tempswafalse\@citex[]}}
\def\citelow{\@cghifalse\@ifnextchar [{\@tempswatrue
	\@citex}{\@tempswafalse\@citex[]}}
\def\@cite#1#2{{$\null^{#1}$\if@tempswa\typeout
	{IJCGA warning: optional citation argument 
	ignored: `#2'} \fi}}

\def\pmb#1{\setbox0=\hbox{#1}
	\kern-.025em\copy0\kern-\wd0
	\kern.05em\copy0\kern-\wd0
	\kern-.025em\raise.0433em\box0}

\def\fnt#1#2{\footnotetext{\kern-.3em
	{$^{\mbox{\scriptsize #1}}$}{#2}}}

\def\fpage#1{\begingroup
\voffset=.3in
\thispagestyle{empty}\begin{table}[b]\centerline{\footnotesize #1}
	\end{table}\endgroup}

\def\runninghead#1#2{\pagestyle{myheadings}
\markboth{{\protect\footnotesize\it{\quad #1}}\hfill}
{\hfill{\protect\footnotesize\it{#2\quad}}}}
\font\tenrm=cmr10
\font\tenit=cmti10 
\font\tenbf=cmbx10
\font\bfit=cmbxti10 at 10pt
\font\ninerm=cmr9

\font\eightrm=cmr8

\def\qed{\hbox{${\vcenter{\vbox{			
   \hrule height 0.4pt\hbox{\vrule width 0.4pt height 6pt
   \kern5pt\vrule width 0.4pt}\hrule height 0.4pt}}}$}}

\begin{document}

\runninghead{The Higgs Boson as a Supersymmetric Partner$\ldots$} 
{The Higgs Boson as a Supersymmetric Partner$\ldots$}

\normalsize\textlineskip
\thispagestyle{empty}
\setcounter{page}{1}

\copyrightheading{}			

\vspace*{0.88truein}

\fpage{1}
\centerline{\bf THE HIGGS AS A SUPERSYMMETRIC PARTNER}
\vspace*{0.035truein}
\centerline{\bf WITH A NEW INTERPRETATION OF YUKAWA COUPLINGS}
\vspace*{0.37truein}
\centerline{\footnotesize ROLAND E. ALLEN}
\vspace*{0.015truein}
\centerline{\footnotesize\it Center for Theoretical Physics, 
Texas A\&M University}
\baselineskip=10pt
\centerline{\footnotesize\it College Station, Texas 77843, USA}


\vspace*{0.21truein}
\abstracts{An unconventional version of supersymmetry leads to the following 
highly testable predictions: (1) The Higgs boson has an R-parity of -1, so 
it can only be produced as one member of a pair of superpartners. (2) The only
superpartners are scalar bosons, so neutralinos etc. do not exist. 
(3) The most likely candidate for cold dark matter is therefore a sneutrino. 
(4) The Higgs and other bosonic superpartners have an unconventional equation of
motion. These predictions are associated with new interpretations of Yukawa
couplings, supersymmetry, gauge fields, and Lorentz invariance.}{}{}

\textlineskip			
\vspace*{12pt}			

The next ten years offer promise for revolutionary new discoveries
in experimental high-energy physics.$^{1}$ For example, there is every reason to
believe that both a Higgs boson$^{2-4}$ and supersymmetry$^{5-9}$ will be 
observed at either the Tevatron or the LHC. In addition, there is strong
evidence for neutrino masses,$^{10-12}$ and terrestrial dark matter
experiments are now beginning to approach discovery potential.$^{13-17}$ 
One hopes that theoretical endeavors in these areas can respond to the 
challenge, and provide {\it testable predictions} together with new 
fundamental insights.

Recently we proposed a theory which does have testable 
predictions.$^{18-20}$ It begins with a Euclidean action

\begin{equation}
S=\int d^{D}x\left[ \frac{1}{2m}\partial ^{M}\Psi ^{\dagger }\partial %
_{M}\Psi -\mu \Psi ^{\dagger }\Psi +\frac{1}{2}b\left( \Psi ^{\dagger }\Psi
\right) ^{2}\right] 
\end{equation}
where 
\begin{equation}
\Psi =\left( 
\begin{array}{c}
z_{1} \\ 
z_{2} \\ 
\vdots  \\ 
z_{N}
\end{array}
\right) \qquad ,\qquad z=\left( 
\begin{array}{c}
z_{b} \\ 
z_{f}
\end{array}
\right) .
\end{equation}
This action has ``natural supersymmetry'',$^{21}$ in the sense that the initial
bosonic fields $z_{b}$ and fermionic fields $z_{f}$ are treated in exactly
the same way. The only difference is that the $z_{b}$ are ordinary complex
numbers whereas the $z_{f}$ are anticommuting Grassmann numbers.

There is an alternative starting point, 
involving a microscopic statistical picture at the Planck scale, 
which leads to the phenomenological action (1).  Anyone interested in 
this picture may consult Sections 2 and 3 of Ref. 20. Here, however, 
we will start closer to experiment, by simply postulating (1).

This action leads to an $SO(10)$ gauge theory,$^{1}$ which contains the
Standard Model of particle physics plus the familiar see-saw mechanism for
small neutrino masses. The present theory also contains Einstein
gravity, as an approximation which holds below the Planck scale. Although
quantization initially involves a Euclidean path integral,$^{18}$ a
consistent canonical formulation is also possible.$^{20}$

In the present theory, Lorentz invariance emerges as a very good
approximation at energies that are low compared to the Planck scale -- or,
to be more precise, for vector bosons at essentially all energies below the
Planck scale and for fermions at energies that are small compared to an
energy $\lambda ^{2}m_{P}c^{2}$. Although there are very sensitive tests of
certain aspects of Lorentz invariance$^{19,22}$ --
such as rotational invariance, locality, microcausality, CPT invariance, and
the requirement that $k^{2}=0$ for massless particles -- these aspects are
unchanged in the present theory.

The present theory thus appears to be consistent with experiment and
observation. The predictions to which it leads, however, are 
highly unconventional. Here we will emphasize the meaning of
these predictions for experiments in the foreseeable future -- specifically
the implications regarding searches for a Higgs boson, supersymmetry, and
dark matter. 

According to (1), the only fundamental fermions $\psi $ are those of an 
$SO(10)$ grand unified theory with three generations, and the only
superpartners are a matching set of scalar bosons $\phi $. (Gauge bosons 
$A_{\mu }$ will be discussed 
below.) These spin $1/2$ fermions and spin $0$ bosons have the same gauge
couplings. For example, one matching set of fields is the electroweak Higgs
doublet 
\begin{equation}
\phi _{h}=\left( 
\begin{array}{c}
\phi ^{+} \\ 
\phi ^{0}
\end{array}
\right) 
\end{equation}
and the charge conjugate $\psi _{\ell }^{c}$ of the left-handed lepton
doublet 
\begin{equation}
\psi _{\ell }=\left( 
\begin{array}{c}
\nu _{L} \\ 
e_{L}
\end{array}
\right) .
\end{equation}
I.e., $\phi _{h}$ is the superpartner of $\psi _{\ell }^{c}$. This means
that it has a lepton number of $-1$, and an R-parity $R=\left( -1\right)
^{3\left( B-L\right) +2s}=-1$, so the physical Higgs boson can only be 
produced in conjunction with another scalar superpartner.

This is, of course, an eminently testable prediction. For example, at the 
time this paper is being written, evidence for a standard Higgs has been
reported at CERN. If this ``discovery'' holds up, the present theory will be
disconfirmed. On the other hand, if a standard Higgs continues not to be
observed, as the favorable range of parameter space is further exhausted, it
may be wise to consider the possibility that the Higgs has properties other
than those predicted by the Standard Model or standard supersymmetry. In the
present theory, the Higgs will behave in somewhat the same way as the
sneutrino of standard SUSY.$^{23,24}$

As in other grand unified theories, lepton number conservation is a very
good approximation at energies that are low compared to the GUT scale of 
$\sim 10^{13}$ TeV. For this and other compelling reasons,$^{20}$ the effective
Yukawa couplings must have the form 
\begin{equation}
\lambda _{eff}=\lambda _{0}\phi _{GUT}^{\dagger }/M
\end{equation}
where $\lambda _{0}$ is dimensionless and $\left\langle \phi _{GUT}^{\dagger
}\phi _{GUT}\right\rangle \sim M^{2} \sim m_{GUT}^{2}$.

Since all superpartners are scalar bosons, the various fermionic sparticles
of standard supersymmetry do not exist. Again, this feature is highly
testable. If a neutralino or chargino is seen at the Tevatron or the LHC,
the present theory will have been disconfirmed.

In the present theory, fermionic partners of gauge bosons are not required
because the gauge bosons are ultimately derived from scalar boson degrees of
freedom. They are, in fact, excitations of a GUT Higgs field which condenses 
in the very early universe, and they correspond to rotations of its 
order parameter $\Psi _{s}$. A program for the future is to study how the 
initial supersymmetry at high energy, with only spin $1/2$ and spin $0$
fields, evolves into a lower energy supersymmetry in which spin $1$ degrees
of freedom replace many of the initial spin $0$ degrees of freedom. One can
then determine in detail whether (i) the Higgs is fully protected from a
quadratic divergence of its self-energy, as in standard
supersymmetry, and (ii) whether the gauge couplings $\alpha _{1}$, $\alpha
_{2}$, and $\alpha _{3}$ still converge to a common value at some energy 
$m_{GUT}\sim 10^{13}$ TeV, as they do in standard SUSY.

Since neutralinos do not exist, the most likely candidate for cold dark
matter is a spin-zero WIMP -- i. e., a sneutrino. This prediction should be
testable in spin-dependent dark matter searches.

We conclude with a fourth unorthodox and testable feature of the present
theory: The Higgs boson, and other scalar superpartners, are predicted to
have an unconventional equation of motion, which will lead to an
unconventional energy-momentum relation and unconventional kinematics.
Specifically, Higgs fields $\phi _{h}$ have the equation of motion 
\begin{equation}
-\left( g^{\mu \nu }D_{\mu }D_{\nu }+i\overline{m}e_{\alpha }^{\mu }\sigma
^{\alpha }D_{\mu }\right) \phi _{h}-\mu _{h}^{2}\phi _{h}+\overline{b}\left(
\phi _{h}^{\dagger }\phi _{h}\right) \phi _{h}=0.
\end{equation}
A standard treatment$^{25}$ then gives a similar equation for the physical Higgs
boson. The extra, first-order term is due to coherent rotations of the 
GUT-scale order parameter $\Psi _{s}$.$^{26}$

In summary, we have presented a series of predictions which are unique to
the present theory and which should be testable in the near future, at
the Tevatron, at the LHC, and in dark-matter experiments: (1) The Higgs
boson has an R-parity of -1, and can only be produced together with another
scalar superpartner. (2)~Neutralinos and other fermionic superpartners do
not exist. (3) The cold dark matter consists of sneutrinos of a new 
kind. (4) The Higgs 
and other scalar superpartners have an unconventional equation of motion
which will lead to unconventional kinematics for energies significantly
above threshold.

\nonumsection{Acknowledgement}
\noindent
This work was supported by the Robert A. Welch Foundation.

\nonumsection{References}

\end{document}